\documentclass{pasj00}
\draft

\begin{document}
\SetRunningHead{R. Kawabata \& S. Mineshige}{Thermally Driven Winds from RIAFs}
\Received{2009/00/00}
\Accepted{2009/00/00}

\title{Thermally Driven Winds from Radiatively Inefficient Accretion Flows}

\author{Ryoji \textsc{Kawabata} and Shin \textsc{Mineshige}} %
\affil{Department of Astronomy, Kyoto University, Kyoto 606-8502}
\email{kawabata@kusastro.kyoto-u.ac.jp}


%

\KeyWords{accretion, accretion disks---black hole physics---hydrodynamics} 

\maketitle

\begin{abstract}

Radiatively inefficient accretion flows (RIAFs) are common feature of low-luminosity accretion flows, including quiescent states of X-ray binaries and low-lunimosity active galactic nuclei.
Thermally driven winds are expected from such hot accretion flows.
By assuming that the flow has self-similarity structure in the radial direction, we solve the vertical structure of the wind and accretion flows simultaneously and evaluate the mass loss rates by wind.
We find that the ratio of the outflow rate to the accretion rate is approximately unity for a viscosity parameter, $\alpha \lesssim 0.1$, despite some uncertainties in the angular momentum and temperature distributions.
That is, the accretion rate in the RIAFs is roughly proportional to the radius.
Moreover, we elucidate the effect of cooling by wind on the underneath accretion flow, finding that this effect could be important for calculating energy spectrum of the RIAF.
Observational implications are briefly discussed in the context of Sgr A*.

\end{abstract}

\section{Introduction}

Accretions onto compact objects are the most energetic processes in the universe, powering X-ray binaries and nuclei of galaxies (Frank et al. 2002; Kato et al. 2008 for reviews).
Observational and theoretical advances have clarified that the accretion flows show various states according to their mass accretion rates (Narayan \& Yi 1995; Abramowicz et al. 1995).
Especially when the mass accretion rate is well below the Eddington value, $\dot{M} \ll \dot{M}_\mathrm{Edd}=L_\mathrm{Edd}/c^2$, the radiative cooling becomes inefficient and significant fraction of gravitational binding energy is converted to the thermal energy of gas (Ichimaru 1977), and such flows are generally called as radiatively inefficient accretion flows (RIAFs) (Narayan 2002; Chap. 9 of Kato et al. 2008).
RIAFs are thought to exist in low-luminosity state of X-ray binaries and nuclei of galaxies (Narayan et al. 1996; Esin et al. 1997; Di Matteo et al. 2003; Yuan et al. 2003).
Advection-dominated accretion flows (ADAFs), vertically integrated solutions of RIAFs with no mass loss nor convection, are known to have self-similar solutions and their vertically averaged Bernoulli functions are positive (Narayan \& Yi 1994; see, however, Nakamura 1998; Abramowicz et al. 2000).
Though the positiveness of Bernoulli function is not a sufficient condition for the outflow, thermally driven disk winds are naturally expected from such hot RIAFs, since significant part of the flows is marginally bound and is in a situation easy to escape to infinity.
The accretion flows lose their mass by the winds as they flow into the central objects.
As a result of mass loss, the accretion rate, $\dot{M}$, is no longer constant in radius $r$.
Its effect is often expressed as $\dot{M} \propto r^s$ with $s$ being constant of order unity (Blandford \& Begelman 1999, hereafter BB99).

There are some observational implications of outflows in accreting systems.
The accretion rates onto neutron stars in soft X-ray transients in quiescent states seem to be smaller than those of white dwarfs in cataclysmic variables with comparable orbital periods, since the typical luminosities in both cases are similar, in spite of their difference of gravitational potential by three orders of magnitude  (Loeb et al. 2001).
This indicates significant outflows in accretion flows, $s \sim 1$.
Outflows from RIAFs also seem to be common in the nuclei of galaxies.
Compact radio/infrared/X-ray source Sgr A*, harboring a supermassive black hole in our Galactic center, 
also has moderate mass loss.
Its mass accretion rate at $r \lesssim 10$ AU ($\sim 100 r_\mathrm{s}$, where $r_\mathrm{s}$ is the Schwarzschild radius) is estimated as $\dot{M} \sim 10^{-7} M_\odot \ \mathrm{yr}^{-1}$ by Faraday rotation measurement (Marrone et al. 2006).
On the other hand, Bondi accretion rate expected by the hot ambient gas around $r \sim 0.04$pc ($ \sim 10^5 r_\mathrm{s}$) is $\dot{M} \sim 10^{-5} M_\odot \ \mathrm{yr}^{-1}$ (Baganoff et al. 2003).
This significant difference between the inner and outer mass accretion rates indicates mass loss from the accretion flow due to the outflow.

The observed broad-band spectra of Sgr A* and soft X-ray transients can also be fitted by RIAF models with moderate outflows, $s \sim 0.3-0.4$, if the direct hating of electrons in RIAF is efficient (Quataert \& Narayan 1999; Yuan et al. 2003).
Moreover some low-luminosity nuclei of galaxies show weaker radiative outputs than that expected from Bondi accretion rates and some of their spectra have large ratios of X-ray to radio luminosities.
This can also be explained by the occurrence of  outflows in RIAFs (Di Matteo et al. 1999, 2000).

Another suggestion comes from the fact that active galactic nuclei show bimodal distributions in their luminosities (Marchesini et al. 2004).
This distribution can be naturally explained by the existence of two states of accretion flows.
One is bright, geometrically thin accretion disks (Shakura \& Sunyaev 1973), while another is low-luminosity RIAFs with significant outflows at small radii surrounded by thin accretion disks at larger radii.
If $s \sim 1$ in the inner RIAF, the radiative luminosity from the inner part becomes, at most, comparable to that from the outer thin disk.
Hence, RIAF surrounded by thin disk is fainter by orders of magnitude than thin disk extending to the black hole, which could make distinct bimodal distributions in the luminosities (Begelman \& Celotti 2004).

Although the mass loss from the accreting flows seems to be common features in RIAFs, consistent evaluation of the mass loss rates by thermally driven winds from underneath accretion flows has been poorly attempted.
The outflow rate can be evaluated from the density and the temperature at the base of the wind.
Since the base of the wind should be within the accretion flow, it is essential to solve the wind and accretion flow simultaneously, although the wind and accretion flow were completely separated in most of previous studies because  of difficulties in solving two-demensional problems. 
Some studies examined the dynamics of the thermally driven wind with constant angular momentum.
They neglected the underneath accretion flow and assumed that the radial gravitational force balances with centrifugal force (e.g. Fukue 1989; Takahara et al. 1989).
The effects of the magnetic field and radiation force on the wind dynamics were also examined by Fukue (2004).
The outflow rate linearly depends on the density at the base of the wind, which was generally specified by unknown parameter (Kusunose 1991).

As independent researches the effects of outflow on the accretion flow were examined by introducing some parameters which characterize the properties of the wind, such as velocity, density, angular momentum, and energy (Kusunose 1991; BB99; Misra \& Taam 2001; Xie \& Yuan 2008).
The dynamics of the wind was not solved in these studies, and so the outflow rates depend on the unknown parameters.

The outflow rate by thermally driven wind and the effect of the outflow on the accretion flow must be evaluated from the same hydrodynamical equations.
This is the motivation of the present study.
The outflow rates by wind from accretion flows have been calculated in the context of accretion disk corona model.
In the corona models, significant outflows are expected (Meyer \& Meyer-Hofmeister 1994, hereafter MMH94; Witt et al. 1997).
We also expect significant outflows from RIAFs, since the flows are hot, like coronal flows.

In this paper we solve the vertical structure of the flow (both of the accretion flow and thermally driven wind) and show that the RIAFs produce non-negligible outflows.
In \S \ref{sec:model}, we present assumptions, basic equations, and numerical procedure of our model.
Then we show the results of outflow rates and the vertical structures of the wind/accretion flows in \S \ref{sec:results}.
We give discussions on the comparison with numerical simulations and on the applications to some observations of outflows in \S \ref{sec:discussion}, and conclusions are given in \S \ref{sec:conclusions}.

\section{Our Model}
\label{sec:model}

We calculate the mass outflow rate by the wind and mass accretion rate by viscosity simultaneously.
We neglect the effects of magnetic fields on the wind dynamics.
Instead, we use the $\alpha$ prescription in viscous stress tensor, which might be the results of turbulence in magnetized plasma (Balbus \& Hawley 1991; Stone et al. 1996).
We consider steady, axisymmetric, and non-radiative flow in a Newtonian potential, $\psi = - GM/\sqrt{r^2+z^2}$, where $M$ is the mass of central object, and solve the Navier-Stocks equations in the cylindrical coordinates ($r, \varphi,z$).
We assume that the physical properties are self-similar in the radial direction, $\dot{M} \propto r^{s}$ and  $v_r, c_\mathrm{s}, \propto r^{-1/2}$, where $s$ is a constant (see also Xie \& Yuan 2008).
Note that the value of $s$ should be determined iteratively for consistency (see below).
According to the self-similar solutions with outflow, the surface density $\Sigma$ is proportional to $r^{s-1/2}$ (BB99).
Then we assume that the power law index of the density $\rho$ in the radial direction is constant regardless of $z$.
Hence we set $\rho \propto r^{s-1/2}$ and $P \propto r^{s-3/2}$ in our model.
Because of these power-law dependences, the radial derivatives of the physical values are on the order of $\partial / \partial r \sim 1/r$.
With these assumptions we can express the physical properties of the flow as functions of vertical coordinate $z$ at a given radius $r$.
We also assume that the outflow by wind is decoupled from the underneath accretion flow at the critical point $z_\mathrm{c}$, where the wind velocity reaches the sound speed, $v_z \simeq c_\mathrm{s}$.
This assumption means that the outflow above the critical point escapes to infinity, and that the viscous heating generated in this outflow never contributes to the heating of underneath accretion flow.
We solve the hydrodynamical equations of the wind and accretion flows along the $z$ axis without separating them, and evaluate the outflow rate of thermally driven wind from this critical point.

\subsection{Basic Equations}

The basic equations in our model basically follow those of MMH94 but with some modifications.
They considered only a coronal flow from the thin disk but did not solve the underneath thin accretion disk.
On the other hand, we are concerned with the RIAF without underneath thin disk, and solve the whole vertical structure of the RIAF and outflow.
As noted above we assume self-similar structure in the radial direction, 
\begin{eqnarray}
\rho \propto r^{s-1/2}, \ P \propto r^{s-3/2}, \ v_r, c_\mathrm{s} \propto r^{-1/2}, \label{eq:similarity}
\end{eqnarray}
where $s$ is constant.
Moreover, following MMH94, we assume the wind geometry, in order to treat the basic equations as one-dimension in the vertical direction.
For the schematic view of our model, see \S \ref{sec:structure}.
 
\subsubsection{Continuity Equation}

The continuity equation in cylindrical coordinates is
\begin{eqnarray}
\frac{\partial}{\partial z} (\rho v_z) + \frac{1}{r} \frac{\partial}{\partial r} (r \rho v_r) = 0.
\end{eqnarray}
Here we assume that the radial mass flux term $r \rho v_r$ is proportional to $r^{s}$ [see equation (\ref{eq:similarity})].
We then get 
\begin{eqnarray}
\frac{\partial}{\partial z} (\rho v_z) + s \frac{\rho v_r}{r} = 0. \label{eq:continuity0}
\end{eqnarray}
At $z \gg r$, the geometry of the stream line becomes spherical, since the pressure gradient becomes important.
This spherical expansion of the stream line is essential for the transonic wind.
We approximately take into account this expansion of the flow by modifying the continuity equation (\ref{eq:continuity0}).
We assume that the cross section of the expanding flux tube in the vertical direction is proportional to $r^2+z^2$ and, following MMH94, we introduce an extended form of the continuity equation,
\begin{eqnarray}
\frac{1}{r^2+z^2} \frac{d}{d z} [ (r^2+z^2) \rho v_z ] + s\frac{\rho v_r}{\sqrt{r^2+z^2}} = 0,  \label{eq:continuity}
\end{eqnarray}
where the first term is the change of the vertical mass flux per unit area and the second term is the net gain of mass by the radial motion.
The denominator of the second term means that the radial scale also expands at large $z$.
Here we ignore the interaction between the winds which come from different radii.
Integrating (\ref{eq:continuity}) from the equatorial plane of the flow at $z=0$ to the critical point at $z=z_\mathrm{c}$, we obtain a condition which $s$ should satisfy, 
\begin{eqnarray}
s=\frac{\left. (r^2+z^2) \rho v_z \right|_{z_\mathrm{c}}}{- \int_{0}^{z_\mathrm{c}} \sqrt{r^2+z^2} \rho v_r dz }. \label{eq:s}
\end{eqnarray}
Here we neglect the vertical mass flux at $z=0$ since, though we do not impose symmetric condition on the equatorial plane, the vertical velocity is so small, $v_z \ll v_r, c_\mathrm{s}$, that the vertical mass flux is negligible  at $z=0$ (see \S \ref{sec:structure}). 
Note that equation (\ref{eq:s}) means that $s$ represent the ratio of the accretion rate ($4 \pi$ times the denominator) to the outflow rate ($4 \pi$ times the numerator), while $s$ was originally defined by the relation $\dot{M} \propto r^s$ (Witt et al. 1997).

\subsubsection{Radial Force Balance}

The equation of radial motion is  
\begin{eqnarray}
\left(v_r \frac{\partial}{\partial r} + v_z \frac{\partial}{\partial z} \right) v_r -  \frac{v_\varphi^2}{r} = - \frac{\partial \psi}{\partial r} - \frac{1}{\rho} \frac{\partial P}{\partial r}. \label{eq:rEoM}
\end{eqnarray}
Near the equatorial plane $z \ll r$, the inertial term is negligible, since $v_r, v_z \ll c_\mathrm{s}$, and the angular velocity becomes  
\begin{eqnarray}
v_\varphi^2 = \frac{G M r^2}{(r^2+z^2)^{3/2}} - \left( \frac{3}{2} - s \right) \frac{P}{\rho},  \label{eq:vp0}
\end{eqnarray}
where we used equation (\ref{eq:similarity}).
The azimuthal velocity becomes small with increasing $z$, and finally the inertia should become important at some large $z$.
Since we do not exactly know the height where the inertia term dominates the centrifugal force $v_\varphi^2/r$ in equation (\ref{eq:rEoM}), we simply assume that the specific angular momentum approaches a small constant value, $v_\varphi = \varepsilon v_\mathrm{K}$, at large $z$, where $\varepsilon$ is a constant parameter much less than unity.
We connect smoothly the two extreme values of the azimuthal velocity, and have
\begin{eqnarray}
\frac{v_\varphi^2}{v_\mathrm{K}^2} = \left( \left[ 1 - \left\{ \left( \frac{3}{2} -s \right) \frac{P}{\rho v_\mathrm{K}^2} + \varepsilon^2 \right\}^n \right] \left( 1+\frac{z^2}{r^2} \right)^{-3n/2} + \left\{ \left( \frac{3}{2} -s \right) \frac{P}{\rho v_\mathrm{K}^2} + \varepsilon^2 \right\}^n \right)^{1/n} \nonumber \\
- \left( \frac{3}{2} -s \right) \frac{P}{\rho v_\mathrm{K}^2}, \label{eq:vp}
\end{eqnarray}
where $n(>0)$ is a parameter which represents how smoothly the two values of angular velocity are connected.
For larger $n$, $v_\varphi$ chenges more steeply.
In this study we assume $n=2$ or 6

\subsubsection{Angular Momentum Conservation}

The equation of angular momentum conservation is
\begin{eqnarray}
\frac{\partial}{\partial z} (\rho v_z l) + \frac{1}{r} \frac{\partial}{\partial r} [ r (\rho v_r l - r t_{r \varphi} ) ] = 0,
\end{eqnarray}
where $l = r v_\varphi$ is specific angular momentum.
We assume that the $r \varphi$-component of viscous stress tensor, $t_{r \varphi}$, is dominant and neglect other components.
We then modify this equation to the same conservation form as continuity equation (\ref{eq:continuity}),
\begin{eqnarray}
\frac{1}{r^2+z^2} \frac{\partial}{\partial z} [ (r^2+z^2) \rho v_z l ] + \left( s + \frac{\partial \ln l}{\partial \ln r} \right) \frac{\rho v_r l - r t_{r \varphi}}{\sqrt{r^2+z^2}} = 0, \label{eq:angmom}
\end{eqnarray}
where we assume $t_{r \varphi} \propto P v_\varphi / v_\mathrm{K}$ (see below for the detailed form of $t_{r \varphi}$) and use equation (\ref{eq:similarity}).
Since $l$ is given by equation (\ref{eq:vp}), we can calculate the radial velocity $v_r$ using continuity equation (\ref{eq:continuity}), 
\begin{eqnarray}
v_r = - \frac{\partial \ln l/\partial \ln z}{\partial \ln l/\partial \ln r} \frac{\sqrt{r^2+z^2}}{z} v_z + \left( \frac{s}{\partial \ln l/\partial \ln r} + 1 \right) \frac{t_{r \varphi}}{\rho v_\varphi}. \label{eq:vr}
\end{eqnarray}
Here the first term is the radial motion by angular momentum conservation and the second term is by viscous angular momentum transport.

\subsubsection{Equation of Vertical Motion}

The equation of vertical motion is
\begin{eqnarray}
v_z \frac{d v_z}{d z} = -\frac{1}{\rho} \frac{dP}{dz} -\frac{GMz}{(r^2+z^2)^{3/2}}, \label{eq:eom}
\end{eqnarray}
where the inertial term $v_r \partial v_z /\partial r$ is neglected since $v_r \ll c_\mathrm{s}$.

\subsubsection{Temperature Distribution of the Flow}
\label{sec:Gamma}

In the model of the accretion flow it is difficult to evaluate the vertical distribution of temperature because of poorly unknown distribution of viscous dissipation and other microphysics (e.g. thermal conduction and/or convective motion).
Hence, when solving the vertical motion of the flow, we assume the polytropic relation in the vertical distribution of the pressure for simplicity, 
\begin{eqnarray}
P \propto \rho^\Gamma, \label{eq:polytropic}
\end{eqnarray}
instead of solving the energy equation, where $\Gamma$ is a constant parameter.
Note that in an adiabatic one-dimensional flow $\Gamma$ is identical to the the ratio of specific heats $\gamma$.
This is not generally the case, however, since the direction of the integration of equations is not same as that of the streamline and since heating by viscous dissipation is present.
If thermal conduction in the vertical direction is efficient in the collisionless flow, isothermal approximation, $\Gamma \simeq 1$, might be valid.
In the disk corona model with thermal conduction, Meyer, Liu, \& Meyer-Hofmeister (2000) show that the temperature distribution is nearly isothermal at the place far from the underneath cool thin disk.
The thermal sound speed, $c_\mathrm{s}^2 \equiv \Gamma P/ \rho$, is a simple algebraic function of $\rho$ in this assumption.
Hence we can determine the critical point and integrate the vertical equations downward from this point. (see \S \ref{sec:numerical})

\subsubsection{Equation of Energy Conservation}

After calculating the vertical structure of the flow using polytropic relation [equation (\ref{eq:polytropic})] for given $s$ and $h_\mathrm{c}$, where $h_\mathrm{c}=c_\mathrm{s}/v_\mathrm{K}$ at the critical point,
we adjust $s$ and $h_\mathrm{c}$ by iteration so as to satisfy the vertically integrated equation of energy conservation. 
When the radiative cooling processes are negligible in the flow, the energy equation is 
\begin{eqnarray}
\frac{\partial}{\partial z} (\rho v_z u) + \frac{1}{r} \frac{\partial}{\partial r} [ r ( \rho v_r u - v_\varphi t_{\varphi r}) ] = 0,
\end{eqnarray}
where $u$ is the total specific energy of gas, 
\begin{eqnarray}
u=\frac{1}{2} (v_r^2 + v_\varphi^2 + v_z^2) + \frac{\gamma}{\gamma-1} \frac{P}{\rho} + \psi.
\end{eqnarray}
We take into account the effect of expansion of the flow and then we have
\begin{eqnarray}
\frac{1}{r^2+z^2} \frac{d}{dz} [ (r^2+z^2) \rho v_z u ] + (s-1) \frac{\rho v_r u -v_\varphi t_{r \varphi}}{\sqrt{r^2+z^2}}=0, \label{eq:energy0}
\end{eqnarray}
where we assume $u \propto r^{-1}$ and $v_\varphi \propto r^{-1/2}$, since most of the mass in the flow concentrates on the equatorial plane and radial energy fluxes at $z > r$ is negligible (\S \ref{sec:structure}).
Hence the vertically integrated equation of energy is
\begin{eqnarray}
\left. (r^2+z^2) \rho v_z u \right |_{z_\mathrm{c}} = (1-s) \int_0^{z_\mathrm{c}} \sqrt{r^2+z^2} (\rho v_r u -v_\varphi t_{r \varphi}) dz, \label{eq:energy}
\end{eqnarray}
and we determine the value of $c_\mathrm{s}$ at the critical point using this equation.
Here we also neglect the vertical energy flux at $z=0$ [see equation (\ref{eq:s})].

\subsubsection{Viscous Stress Tensor}

Since the vertical distribution of viscous stress tensor is poorly unknown, we simply assume that the kinematic viscosity is proportional to the product of sound speed and pressure scale height, $\nu \sim \alpha c_\mathrm{s} H \sim \alpha c_\mathrm{s}^2 /v_\mathrm{K}$ and that $r \varphi $ component is dominant.
We thus set 
\begin{eqnarray}
t_{r \varphi} = - \alpha P \frac{v_\varphi}{v_\mathrm{K}},
\end{eqnarray}
where $\alpha$ is viscosity parameter (Shakura \& Sunnyaev 1973).
Even if this $\alpha$ viscosity prescription might not be valid at large $z$, our results will hardly be affected, since the energy and angular momentum transports by viscosity are not appreciable at large $z$ (see \S \ref{sec:structure}).

\subsection{Numerical Calculations}
\label{sec:numerical}

In this section we explain the procedure of finding the solutions satisfying the equation of dynamics and energy conservation.

We calculate the vertical structure of the polytropic wind/accretion flow at a given $r$.
When we use $x \equiv z/r$ and velocities normalized by $v_\mathrm{K}$, the equations have only $x$ as an independent variable, and do not explicitly depend on $r$ because of self-similarity.
Moreover, the equations do not depend on the absolute value of the density so that we can normalize the physical values by the density on the equatorial plane, $\rho_0$.
We have wind equations by using equations (\ref{eq:continuity}), (\ref{eq:vp}), (\ref{eq:vr}),  (\ref{eq:eom}), and (\ref{eq:polytropic}),
\begin{eqnarray}
D(x, v_z, c_\mathrm{s}) \frac{d v_z}{dx} = N(x, v_z, c_\mathrm{s}), 
\end{eqnarray}
where $D$ and $N$ are function of $x$, $v_z$ and $c_\mathrm{s}$. 
Solving equation $D=N=0$, we find the critical point, $x_\mathrm{c}$ for given $s$ and $h_\mathrm{c}$, where  subscript c denotes the value at the critical point and $h \equiv c_\mathrm{s}/v_\mathrm{K}$.
By solving the wind equation of the flow, we have the distribution of the mass and energy fluxes in the radial and vertical directions.

We then evaluate the vertically integrated fluxes of mass and energy.
Adjusting values of $s$ and $h_\mathrm{c}$, we finally find consistent solutions satisfying the constraints (\ref{eq:s}) and (\ref{eq:energy}).

\section{Results}
\label{sec:results}

\subsection{Consistent Solutions}

\begin{figure}
  \begin{center}
    \FigureFile(80mm,80mm){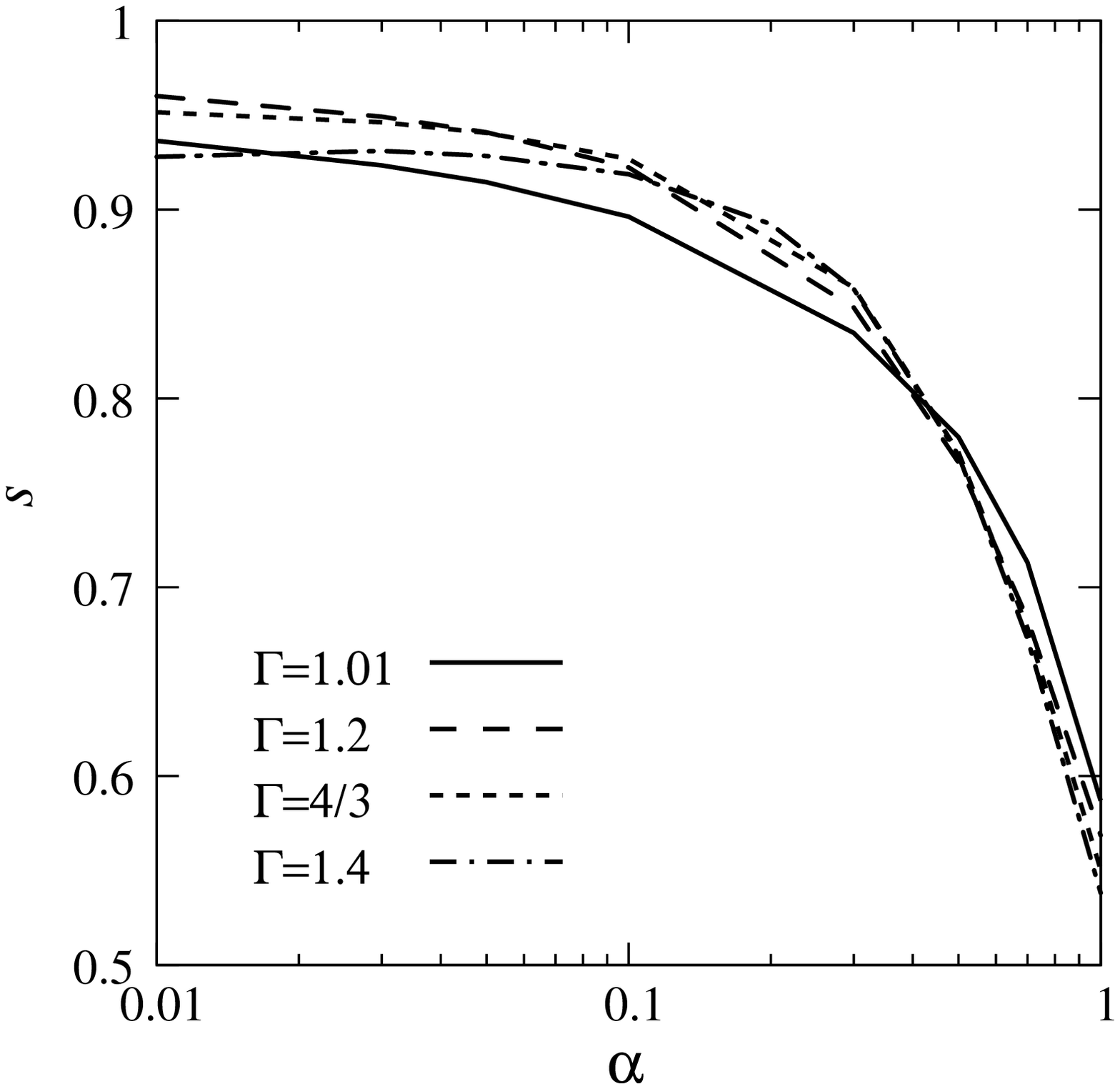}
    \FigureFile(80mm,80mm){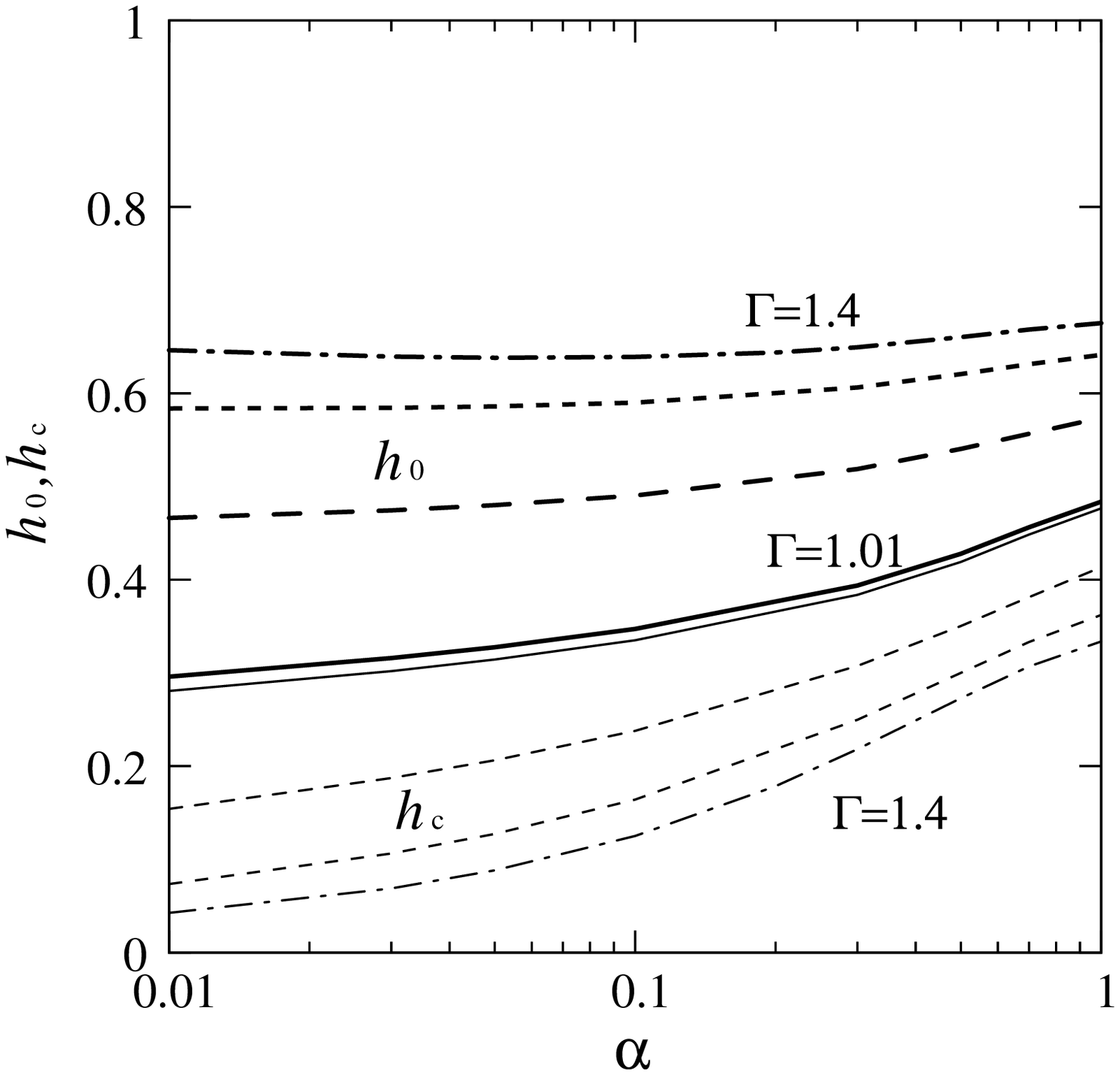}
  \end{center}
  \caption{
Consistent solutions $s$ (left panel) and $h_0$ and $h_\mathrm{c}$(right panel) as functions of viscous parameter $\alpha$ for $\Gamma=1.01$ (solid \ curves), $\Gamma=1.2$ (long-dashed curves), $\Gamma=4/3$ (short-dashed curves), and $\Gamma=1.4$ (dash-dotted curves).
Other parameters are $n=2$, $\varepsilon=0.2$, and $\gamma=5/3$.
 }
 \label{fig:solution1}
\end{figure}

Figure \ref{fig:solution1} shows the ratio of mass outflow rate to the accretion rate, $s$ (left panel), and thermal sound speed on the equator, $h_0$, and that at the critical point, $h_\mathrm{c}$ (right panel), for various $\Gamma$.
The ratio, $s$, decreases when viscous parameter, $\alpha$, increases, since the mass accretion rate increases, while the outflow rate not.
For a given $\alpha$, $s$ does not strongly depend on the temperature distribution $\Gamma$ and for reasonable values of $\alpha$ (e.g. $\alpha \lesssim 0.1$), we obtain $s \sim 0.9$.
This means that significant amount of gas is blown away by the thermally driven wind and mass accretion rate is almost proportional to $r$.
On the other hand, the sound speed, $h_0$, decreases, when $\Gamma$ approaches to unity or when $\alpha$ decreases, since the energy loss by the wind becomes large.
We cannot find consistent solutions for large $\Gamma$, probably because there is no transonic wind solution for $\Gamma \sim 5/3$ (e.g. Holzer \& Axford 1970).

\begin{figure}
  \begin{center}
    \FigureFile(80mm,80mm){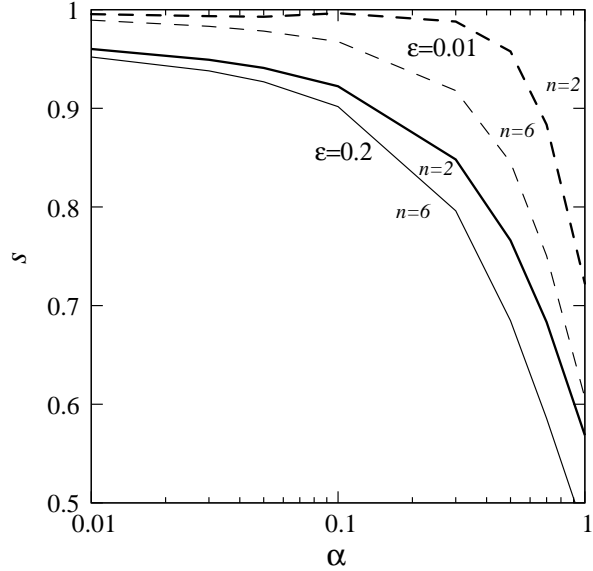}
  \end{center}
  \caption{Same as Fig. \ref{fig:solution1} (left) but for various $n$ and $\varepsilon$ in the cases of $\gamma=5/3$ and $\Gamma=1.2$. 
This figure shows the dependence of $s$ on the angular momentum distributions; $\varepsilon=0.2$ (solid curves) and $\varepsilon=0.01$ (dashed curves), and $n=2$ (thick curves) and $n=6$ (thin curves). }
\label{fig:solution2}
\end{figure}

We also study the dependence of $s$ on the distribution of specific angular momentum $l=r v_\varphi$, i.e., the parameters $n$ and $\varepsilon$ [see equation (\ref{eq:vp})].
Figure \ref{fig:solution2} is same as Figure \ref{fig:solution1} (left) but for various $n$ and $\varepsilon$.
The outflow rates slightly depend on the angular momentum distributions; the larger $n$ or $\varepsilon$ is, the smaller $s$ becomes, though $s$ is larger than $\sim 0.9$ for reasonable values of $\alpha (\lesssim 0.1)$.
We can see that the sound speed of the flow does not depend on the angular momentum distribution.
Hence our results of $s \sim 1$ for $\alpha(\lesssim0.1)$ are robust, regardless of the angular momentum distributions.

\subsection{Vertical Structure of Wind/Accretion Flow}
\label{sec:structure}

\begin{figure}
  \begin{center}
    \FigureFile(80mm,80mm){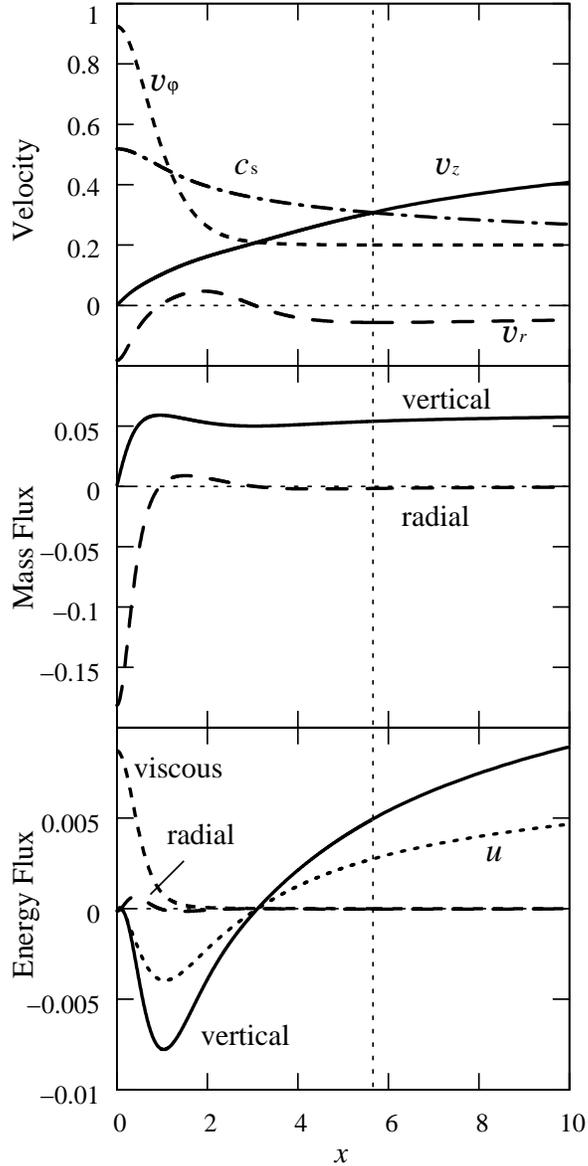}
  \end{center}
  \caption{The vertical structures of wind/accretion flows. 
The parameters are $\gamma=5/3$, $\Gamma=1.2$, $\varepsilon=0.2$, and $n=2$. 
The position of the critical point, $x_\mathrm{c}\simeq5.7$, is shown by vertical dotted line.
Upper panel:
The velocity distribution of the flow, $v_z$(solid curve), $v_r$(long-dashed curve), and $v_\varphi$(short-dashed curve).
The velocities are normalized by $v_\mathrm{K}$.
Sound speed $c_\mathrm{s}$ is also plotted (dash-dotted curve).
Middle panel:
The mass flux distributions normalized by $\rho_0 v_\mathrm{K}$.
Vertical mass flux (solid curve) and radial mass flux (dashed curve) are shown.
The ratio of the outflow rate to the accretion rate is $s=0.919$.
Lower panel:
The distribution of energy flux normalized by $\rho_0 v_\mathrm{K}^3$.
We show the vertical flux (solid curve), radial flux (long-dashed curve), and energy transport by viscosity (short-dashed curve) respectively.
The total energy $0.03 u /v_\mathrm{K}^2$ is also plotted for reference (dotted curve) }
\label{fig:structure}
\end{figure}

By solving the vertical equations of the flow, we obtain the vertical structure of wind and accretion flow simultaneously.
Figure \ref{fig:structure} (upper) shows the velocity distributions in the flow.
Radial velocity $v_r$ is negative (i.e., accretion) at $z \ll  r$ where the angular momentum transport by viscosity is effective.
While, at $z>r$, $v_r$ becomes positive since $v_z$ becomes so large that the first term on the right-hand side of equation (\ref{eq:vr}) (representing radial motion by angular momentum conservation) dominates over the second term (representing radial motion by viscosity).
Though the radial velocity depends on the distribution of angular momentum at $z>r$, large fraction of the mass accretes at $z \ll r$ where the density is high (see below).
At $z=0$ the vertical motion is so small, $v_z \ll c_\mathrm{s}, v_r$, that we can neglect the vertical fluxes on the equator.
This validates our assumption that mass and energy fluxes in the vertical direction at $z=0$ are negligible.
We also find that $v_z$ at $z=0$ is smaller for smaller $\Gamma$.
This dependence might imply that we should consider the vertical distribution of the temperature carefully to satisfy the exact symmetry condition $v_z=0$ at $z=0$.

Note that we plot the distributions of the physical quantities over the critical point using polytropic relation $P \propto \rho^\Gamma$, even though we do not consider the energy balance at $x>x_\mathrm{c}$.

Figure \ref{fig:structure} (middle) shows the mass fluxes of the flow as functions of $z$; vertical mass flux $(1+x^2) \rho v_z$ and radial mass flux $\sqrt{1+x^2} \rho v_r$ .
Mass accretion by viscosity occurs mainly at $z \ll r$, where $\alpha$ viscosity prescription could be valid.
Small fraction of the mass flows outward in the radial direction because of angular momentum conservation, and its fraction is smaller for larger $\alpha$.
At large $z \gg r$, vertical mass flux is constant, since the radial mass flux becomes negligible in the low density region.

We also show the vertical distributions of the energy fluxes in Figure \ref{fig:structure} (lower); vertical energy flux $(1+x^2) \rho v_z u$, radial energy flux $(1-s) \sqrt{1+x^2} \rho v_r u$, and energy flux by viscosity $- (1-s) \sqrt{1+x^2} v_\varphi t_{r \varphi}$.
Note that we assume the temperature distribution in the flow as polytropic and only consider the vertically integrated energy equations (\ref{eq:energy}), instead of solving the energy equation (\ref{eq:energy0}) at each $z$.
We can see that the distribution of the total specific energy $u$ and also of radial energy flux depends on $\Gamma$.
For large $\Gamma$, $u$ at $z \ll r$ is positive and for $\Gamma \sim 1$ it is negative, while $u_\mathrm{c}$ is always positive.
Nevertheless the radial energy flux and viscous energy flux are negligible at $z >r$ for any $\Gamma$.

\begin{figure}
  \begin{center}
    \FigureFile(80mm,80mm){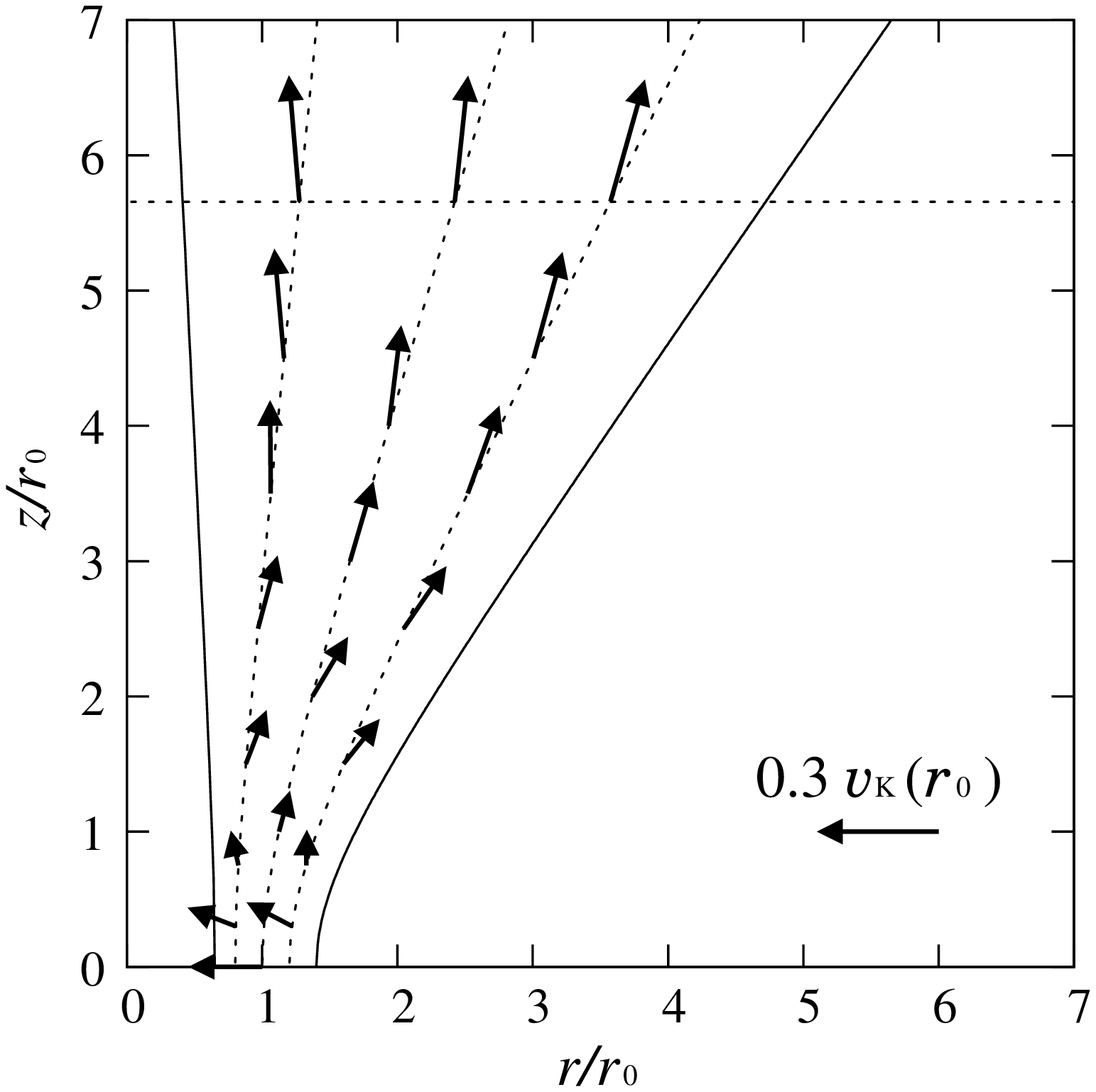}
  \end{center}
  \caption{Schematic view of wind and accretion flow around $r=r_0$ in $r$-$z$ plane.
The velocities are same as in Fig. 3 and the position of critical points ($z/r_0=5.7$) is shown by horizontal dotted line.}
\label{fig:geometry}
\end{figure}

In figure \ref{fig:geometry},  we show a schematic view of wind and accretion flow with velocity fields.
The wind is launched at $r \sim r_0$ and the cross section of the flow is proportional to $z^2$ at large $z$.
The direction of velocity is radial at the equatorial plane and almost spherical at large $z$.

\section{Discussion}
\label{sec:discussion}

\subsection{The Dependence of $s$ on $\alpha$}

We evaluated the mass loss rate by wind from RIAF by assuming that the flow has the self-similar structure in the radial direction.
Though there are some uncertainties in the angular momentum and temperature distributions, we can expect significant mass loss from RIAF, especially when $\alpha$ is small.
We can see the dependence of $s$ and $h_0$ on $\alpha$ as follows.
The mass accretion rate is roughly proportional to $\alpha$, while the outflow rate is determined only by $h_0$.
If $\alpha$ is sufficiently large (e.g. $\alpha \sim 1$) the mass accretion rate is large compared with the outflow rate.
In this case the cooling effect by the wind is not important and the temperature of the flow on the equator is nearly the same as that of classical ADAF with no outflow.
When $\alpha<1$, conversely, the outflow rate is comparable with the accretion rate, and then the flow is effectively cooled by the wind, and hence $h_0$ decreases.

When $\alpha$ decreases further and $s$ approaches unity, the right-hand side of equation (\ref{eq:energy}) vanishes.
Therefore, the ratio of the wind cooling term [the left-hand side of equation (\ref{eq:energy})] to the radial energy transport term [the integral in the right-hand side of equation (\ref{eq:energy})] becomes large, in proportion to $1/(1-s)$.
Then, for $\alpha \ll 1$, the temperature of the flow decreases by the wind cooling and settles to the equilibrium value, for which $s$ is almost unity, but cannot exceed unity, since the left hand side of the equation (\ref{eq:energy}) is always positive.  

\subsection{Comparison with Numerical Simulations}

There are some numerical simulations of hydrodynamic and magnetohydrodynamic (MHD) accretion flows which show the outflows.
We found that $s$ is larger for smaller $\alpha$, which agrees with the analytical one-dimensional model by Misra \& Taam (2001).
However these results are opposite to the results of two-dimensional hydrodynamical simulations by Igumenshchev \& Abramowicz (2000).
The effect of radial convective motion might explain this discrepancy.
It is known that the non-radiating ADAF becomes convectively unstable for low $\alpha$ (Narayan \& Yi 1994).
Igumenshchev \& Abramowicz (2000) showed that in the accretion flow with low $\alpha$ the convective motion becomes prominent and transports the angular momentum inward (convection dominated accretion flows: CDAFs; Igumenshchev et al. 1996; Quataert \& Gruzinov 2000).
MHD simulations by Machida, Matsumoto, \& Mineshige (2001) also show convection.
This radial convective motion, which is not considered in our model, might suppress the wind outflow.

On the other hand three-dimesional MHD simulations show that outward angular momentum transport  occurs by magnetic processes (Armitage 1998; Machida \& Matsumoto 2003).
Hawley \& Balbus (2002) showed that in radiatively inefficient MHD flow significant fraction of the gas is blown away by wind, and the mass accretion rate is proportional to the radius, $s \sim 1$, which is consistent with our results.

\subsection{Observations of the Wind}

Our result that shows significant outflow ($s \sim 1$) in RIAF with moderate $\alpha$  is consistent with the implied mass loss in Galactic X-ray binaries (Loeb et al. 2001).
However such strong wind seems to be inconsistent with the moderate wind suggested to exist in Sgr A*, since strong mass loss cannot explain the observed spectra of Sgr A* (Quataert \& Narayan 1999; Yuan et al. 2003).
There are two possibilities to avoid the significant mass loss in Sgr A*.
First the wind in the flow may be intrinsically small because of large $\alpha$.
Second the RIAF might exist only at the inner radius.
If the supplied gas to Sgr A* has small angular momentum, the flow behaves as a Bondi accretion at large radius (Proga \& Begelman 2003; Cuadra et al. 2006; Mo\'{s}cibrodzka et al. 2006).
Within the circularization radius, for example $r_\mathrm{circ} \sim 100 r_\mathrm{s}$, the flow could become centrifugally supported RIAF and the mass loss starts.

We also show that the sound speed (i.e., temperature of the gas) on the equator, $h_0$, strongly depends on the temperature distribution, $\Gamma$. 
For example the temperature difference between the case with $\Gamma=1.01$ and that with $\Gamma=1.4$ is about factor of three when we consider monoatomic gas, $P / \rho \propto T$ (see Fig. \ref{fig:solution1} right).
This temperature difference could be important for evaluating the spectral energy distribution from RIAFs, since the emissivities strongly depend on the electron temperature, $T_\mathrm{e}$.
In the case of thermal synchrotron emission, the luminosity is proportional to $T_\mathrm{e}^7$ (Mahadevan 1997).
Hence the luminosity could be lowered by three orders of magnitude for $\Gamma \simeq 1$ than for large $\Gamma$, as long as we assume $T_e \propto T$ for a given mass accretion rate.
The electron temperature in RIAF should be evaluated by the electron energy equation including the Coulomb coupling with protons, adiabatic compression, and unknown direct heating by the magnetic field, which is beyond the scope of our paper.
To summarize, if the energy loss by wind is significant, the flow is cooled and radiative spectra might be different from that of classical ADAFs.

Hence it is important to take into account the vertical temperature distribution when calculating the energy spectra.
We will calculate the vertical structure of energy transport including radiative cooling and thermal conduction in two-temperature plasma in future work.

\section{Conclusions}
\label{sec:conclusions}

We evaluated the outflow rates from RIAFs by solving the vertical structure of the wind and accretion flows.
Our results show that strong wind from RIAF is naturally expected and the ratio of the outflow rate to the accretion rate is approximately unity, which mean that the accretion rate in RIAFs is proportional to the radius, $\dot{M} \propto r$.
Moreover the underneath accreting flow is strongly affected not only by mass loss but also by energy loss.
Consequently RIAF with strong wind could have lower temperature by a factor of a few than that of classical ADAFs, which could make significant differences in the observed flux, by three orders of magnitude at largest. 

\bigskip

The authors are grateful to J. Fukue for useful discussions and comments.
This work is supported in part by the Grant-in-Aid of MEXT (19340044, SM) and by the Grant-in-Aid for the global COE programs on "The Next Generation of Physics, Spun from Diversity and Emergence"
from MEXT.

\end{document}